\documentclass{ws-ijmpe}

\newcommand{\beq}{\begin{equation}}
\newcommand{\eeq}{\end{equation}}
\newcommand{\bea}{\begin{eqnarray}}
\newcommand{\eea}{\end{eqnarray}}

\begin{document}

\catchline{}{}{}{}{}

\title{Quantum fluctuations and stability of tetrahedral deformations in atomic nuclei.}

\author{K. Zberecki, P. Magierski}
\address{Faculty of Physics, Warsaw University of Technology, ul. Koszykowa 75, 00-662 Warsaw, Poland}
\author{P.-H. Heenen}
\address{Service de Physique Nucl\'{e}aire Th\'{e}orique, U.L.B - C.P. 229, B 1050 Brussels, Belgium}
\author{N. Schunck}
\address{Institute of Theoretical Physics, ul. Hoza 69, 00-631 Warsaw, Poland}
\address{University of Surrey, Guildford GU2 7XH, UK}

\maketitle

\begin{history}
\received{(received date)}
\revised{(revised date)}
\end{history}

\begin{abstract}
The possible existence of stable axial octupole and tetrahedral
deformations is investigated in $^{80}$Zr and $^{98}$Zr. HFBCS
calculations with parity projection have been performed for
various parametrizations of the Skyrme energy functional. The
correlation and excitation energies of negative parity states
associated with shape fluctuations have been obtained using the
generator coordinate method (GCM). The results indicate that in
these nuclei both the axial octupole and tetrahedral deformations
are of dynamic character and possess similar characteristics.
Various Skyrme forces give consistent results as a function of
these two octupole degrees of freedom both at the mean-field level
as well as for configuration mixing calculations.
\end{abstract}

\section{Introduction}

It has been recently conjectured that many nuclei throughout the periodic table
possess a tetrahedral deformation in their ground- or low-lying isomeric state.
This type of deformation is realized mainly through the non-zero intrinsic octupole
moment $Q_{32} \propto r^{3}(Y_{32}+Y_{3-2})$ accompanied by vanishing quadrupole
deformation.
The first study pointing at the importance of the tetrahedral degree of freedom
in many-fermion systems has been reported in Ref.~\cite{hmx}.
Recently more realistic approaches based
on the microscopic-macroscopic model with Woods-Saxon Hamiltonian revealed
that several nuclei may possess stable tetrahedral deformation.
These nuclei are believed to form islands on the nuclear chart around
"tetrahedral magic" numbers of neutrons and protons:
$16, 20, 32, 40, 56-58, 70, 90-94$\cite{ld,dgs,dud}.
The enhanced susceptibility towards the tetrahedral deformation
is associated with the symmetry of pyramid-like shapes.
It leads to the appearance of two- and four-fold  degeneracies
in the single-particle spectrum and consequently generates large shell
effects stabilizing this deformed configuration \cite{dgs,ld,dud,dgs2}.

On the other hand the energy difference between the spherical and tetrahedral
configurations is a sensitive function of the pairing strength.
Consequently a stable tetrahedral deformation
is a result of a delicate balance between shell effects and pairing correlations.
Indeed it was shown that for two
tetrahedral magic nuclei: $^{80}$Zr and $^{98}$Zr, the
energy difference between spherical and tetrahedral configurations does
not exceed  $1$ MeV. Consequently
quantum fluctuations beyond the mean-field play a significant role \cite{zber}.
Moreover these nuclei also exhibit other types of octupole deformations: the
coupling between the neutron $d_{5/2}$ and $h_{11/2}$ orbitals and
the proton $p_{3/2}$ and $g_{9/2}$ orbitals leads to both axial
and non-axial octupole correlations \cite{bn,zber}. Moreover octupole
deformations are in competition with the quadrupole mode.

In the present article we analyze the influence of various
parametrizations of the Skyrme interaction and pairing
strengths on the existence of
tetrahedral deformation in $^{80}$Zr and $^{98}$Zr.
In the first section the competition between axial octupole and
tetrahedral shapes at the mean-field level is discussed.
In the next section the nuclear dynamics beyond the mean-field
is investigated. This study is directed towards
the determination of the correlation energies and
excitation energies of negative parity states.

\section{Hartree-Fock + BCS approach and the parity projection.}

The Hartree-Fock + BCS (HFBCS) method has been applied to obtain the
energy of a nucleus as function of octupole degrees of freedom.
The Hartree-Fock equations have been solved on a 3-dimensional mesh in
coordinate space.
The details of the calculations can be found in Refs.~\cite{bfh05,Hee91,zber}.
The pairing interaction has been treated in the BCS
approximation including the Lipkin-Nogami (LN) correction~\cite{ln}. A
zero-range density-dependent pairing interaction has been used:
\begin{equation}
V_{pair}=\frac{1}{2}g_{i}( 1 -P_{\sigma} )
\delta({\bf r}-{\bf r'})\left ( 1 - \frac{\rho({\bf r})}{\rho_{0}}
\right ),
\end{equation}
where $i=n,p$ for neutrons and protons, respectively. As in
previous applications, we set $\rho_{0} = 0.16 \text{fm}^{-3}$.
The strength of the pairing force has been adjusted to "experimental"
pairing gaps, extracted from the odd-even mass staggering using a
three-point filter from Ref.~\cite{dmn}.

The behavior of the HFBCS energy as function
of the axial octupole and tetrahedral degrees of freedom
qualitatively agrees with the results presented in Ref.~\cite{yam,tak,polb}.
The existence of axial and non-axial octupole minima is
a sensitive function of the pairing strength. The minima are not
well pronounced and vanish when pairing is increased \cite{zber}.
However the HFBCS approach does not conserve neither
parity nor particle number. Namely, the HFBCS
solutions represents a mixture of positive and negative
parity states.

\begin{figure}
  \begin{center}
    \begin{tabular}{cc}
      \resizebox{60mm}{!}{\includegraphics[angle=270]{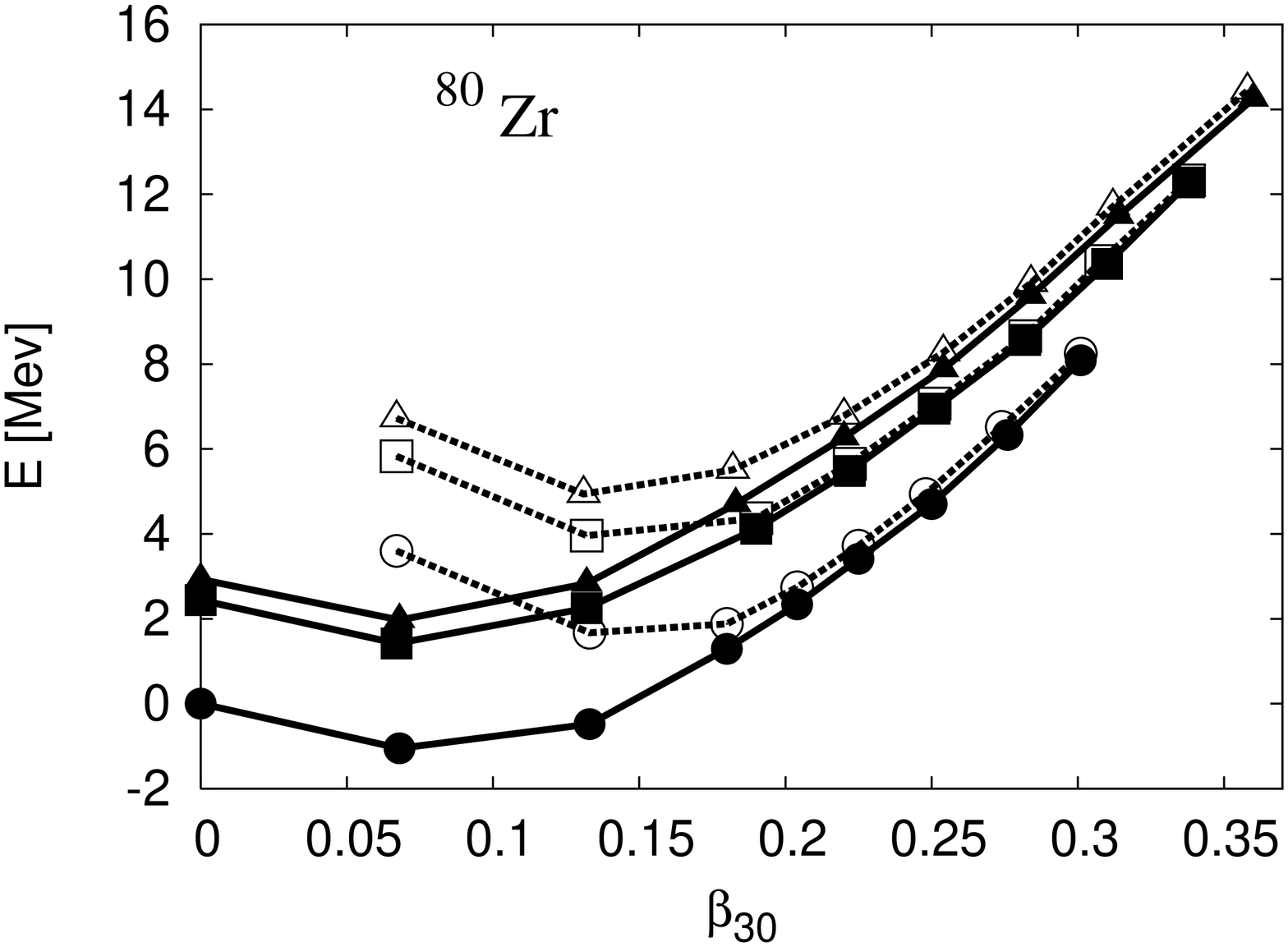}} &
      \resizebox{60mm}{!}{\includegraphics[angle=270]{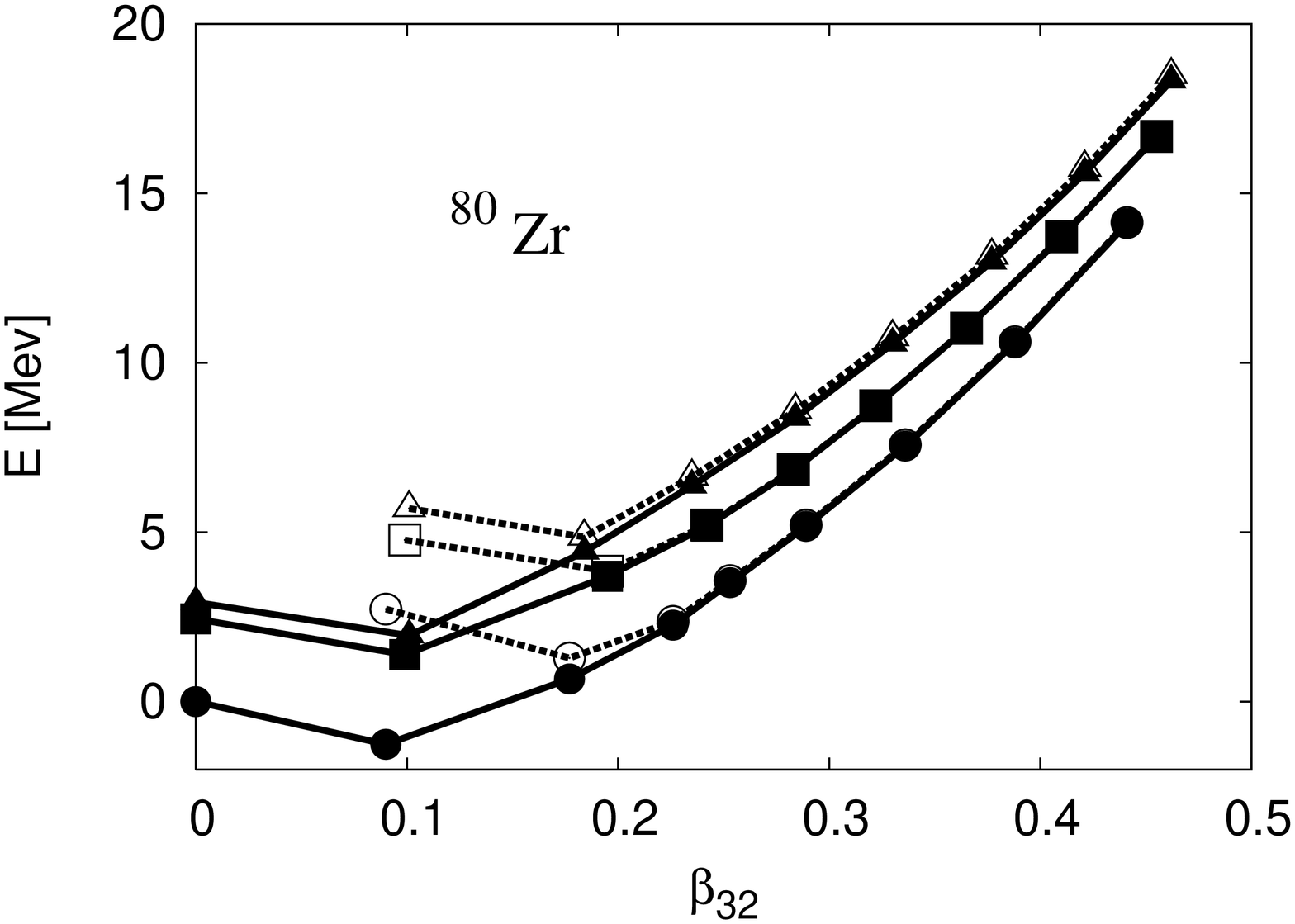}} \\
      \resizebox{60mm}{!}{\includegraphics[angle=270]{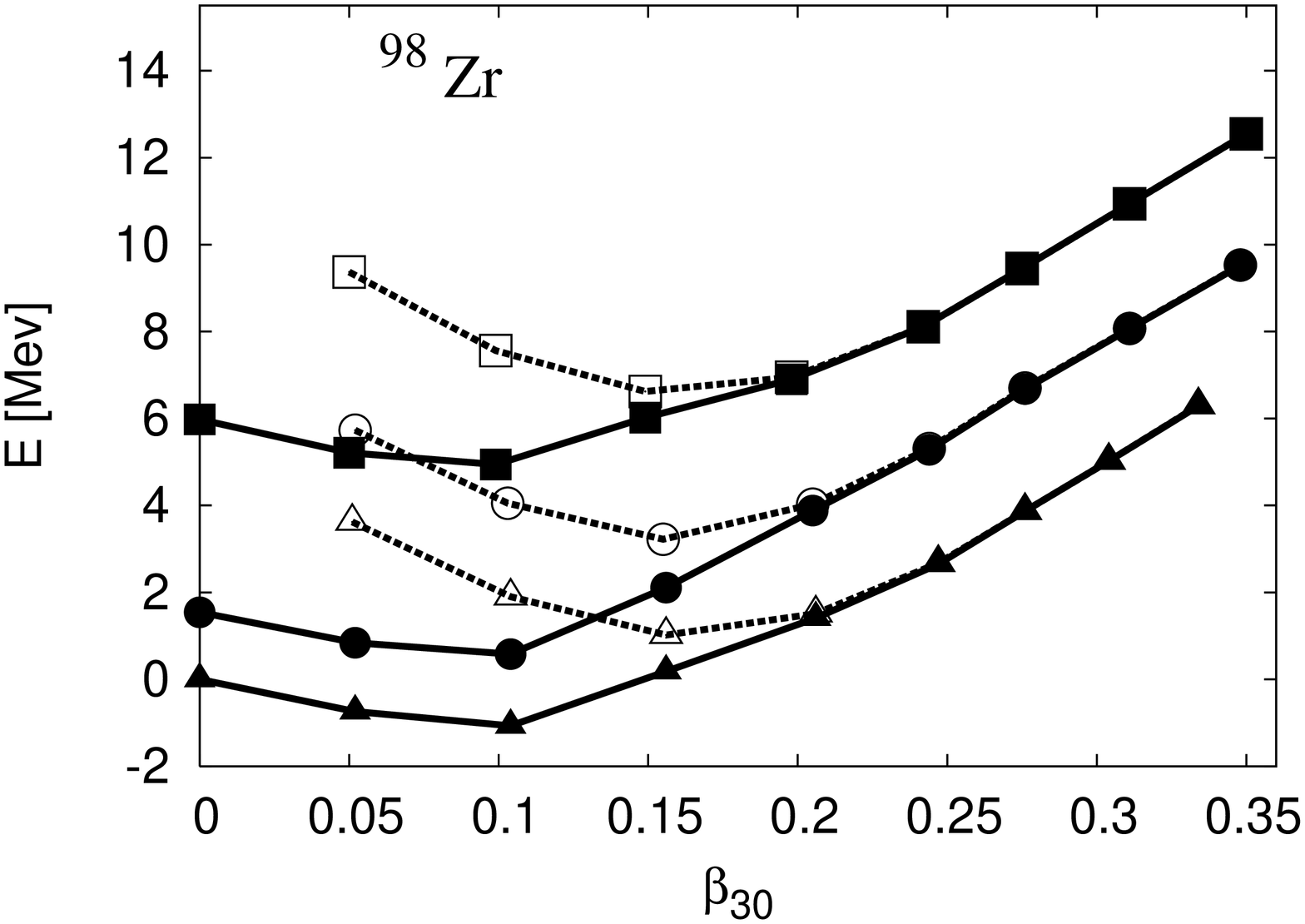}} &
      \resizebox{60mm}{!}{\includegraphics[angle=270]{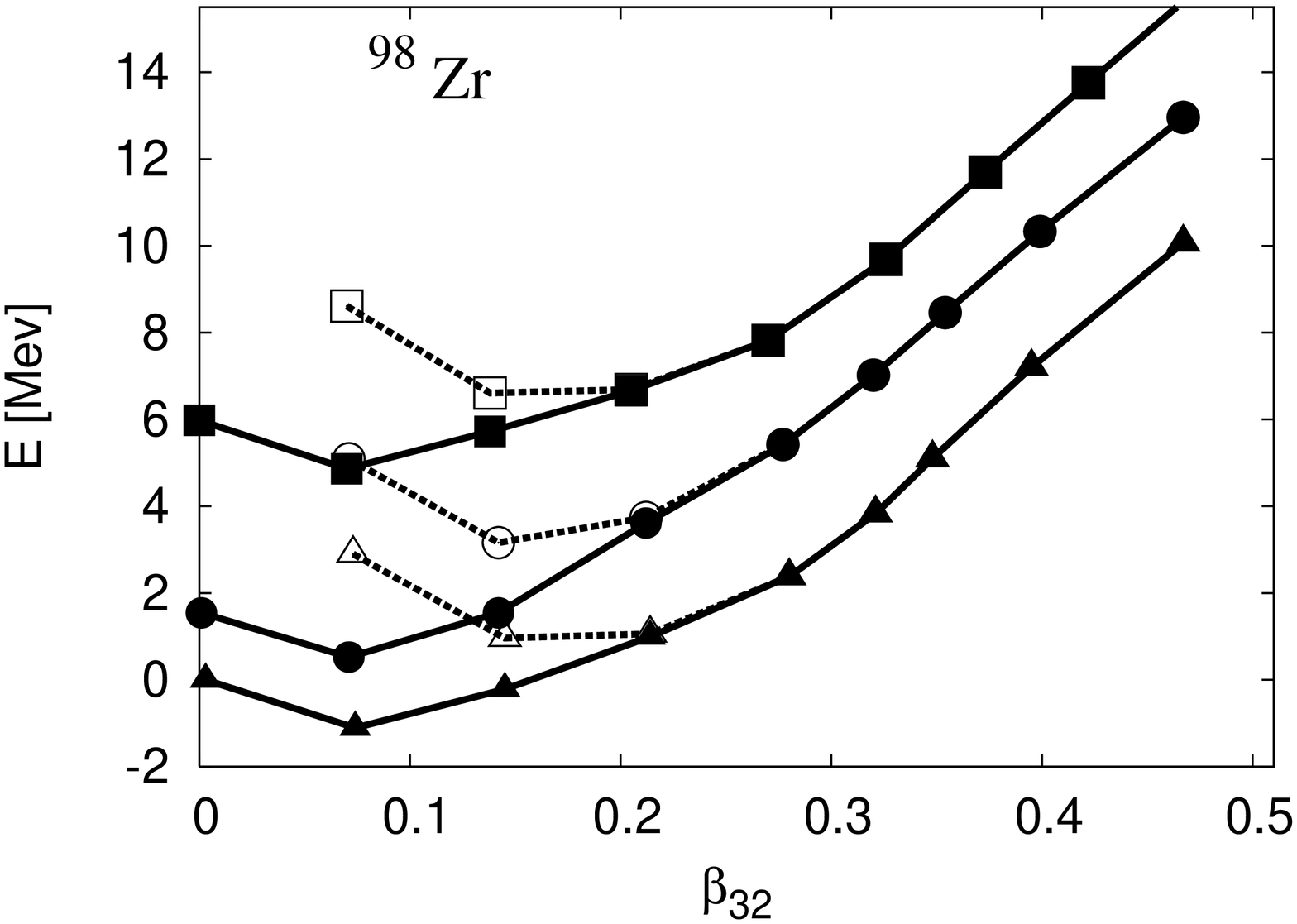}} \\
    \end{tabular}
    \caption{\label{fig1} The parity-projected mean-field energy as function of
shape parameters $\beta_{3\mu}$ (see text for definition) is shown for
three parametrizations of the Skyrme force. The two figures at the left-hand side show the
energy as function of the axial octupole shape parameter $\beta_{30}$,
whereas the figures at the right-hand side show the
energy as function of $\beta_{32}$.
The solid lines with filled symbols denote positive parity solutions and
the dotted lines with open symbols denote negative parity solutions.
Circles, triangles and squares correspond to the Skyrme
parametrization: Sly4, SkM$^{*}$ and SIII, respectively.
The energies are shown relative to the lowest energy
of the spherical configuration.}
  \end{center}
\end{figure}

The restoration of broken symmetries has been performed through
the projection on a definite parity and particle number. The
corrected energies for $N$ neutrons, $Z$ protons and both parities
are obtained by exact projection:
\beq
E(N,Z,\beta_{3\mu})_{\pm} = \frac{\langle
\phi(\beta_{3\mu})|\hat{H}\hat{P}_{(\pm, N,
Z)}|\phi(\beta_{3\mu})\rangle} {\langle
\phi(\beta_{3\mu})|\hat{P}_{(\pm, N,
Z)}|\phi(\beta_{3\mu})\rangle} ,
\eeq
where $|\phi(\beta_{3\mu})\rangle$ are HFBCS wave functions
generated with the constraint $\langle
\phi(\beta_{3\mu})|\hat{Q}_{3\mu}|\phi(\beta_{3\mu})\rangle=C_{\mu}\beta_{3\mu}$.
The shape parameters $\beta_{3\mu}$ are related to the octupole moments
through the relation: $\beta_{30}=\langle Q_{30}\rangle /C_{0} $,
$\beta_{32}=\langle Q_{32}\rangle/C_{2} $, where
$C_{0}=\displaystyle{\frac{3}{4 \pi}} A^{2} r_{0}^{3},
C_{2}=C_{0}/\sqrt{2}$ with $r_{0}=1.2 fm$.
The operator $\hat{P}_{(\pm,N,Z)}$ is the product of operators
projecting  on $\pi = \pm 1$ parity  and on $N$ neutrons and $Z$
protons.

The parity-projected energies for three parametrizations
of the Skyrme force are shown in the Fig.~\ref{fig1}.
One may notice that all forces give qualitatively
the same dependence (apart from a trivial energy shift)
as function of the axial octupole and tetrahedral degree of freedom.
As usual after parity
restoration~\cite{shb93a} the energy minima for positive parity
states are shifted towards smaller octupole deformations compared
to the HFBCS minima, while the negative parity
states have larger deformations. For both nuclei the energy minima
for positive parity correspond to very similar $\beta_{30}$ and
$\beta_{32}$ values. For the negative parity curve, $\beta_{32}$
is systematically larger than $\beta_{30}$ in the minimum.

\begin{figure}
  \begin{center}
    \begin{tabular}{cc}
      \resizebox{60mm}{!}{\includegraphics[angle=270]{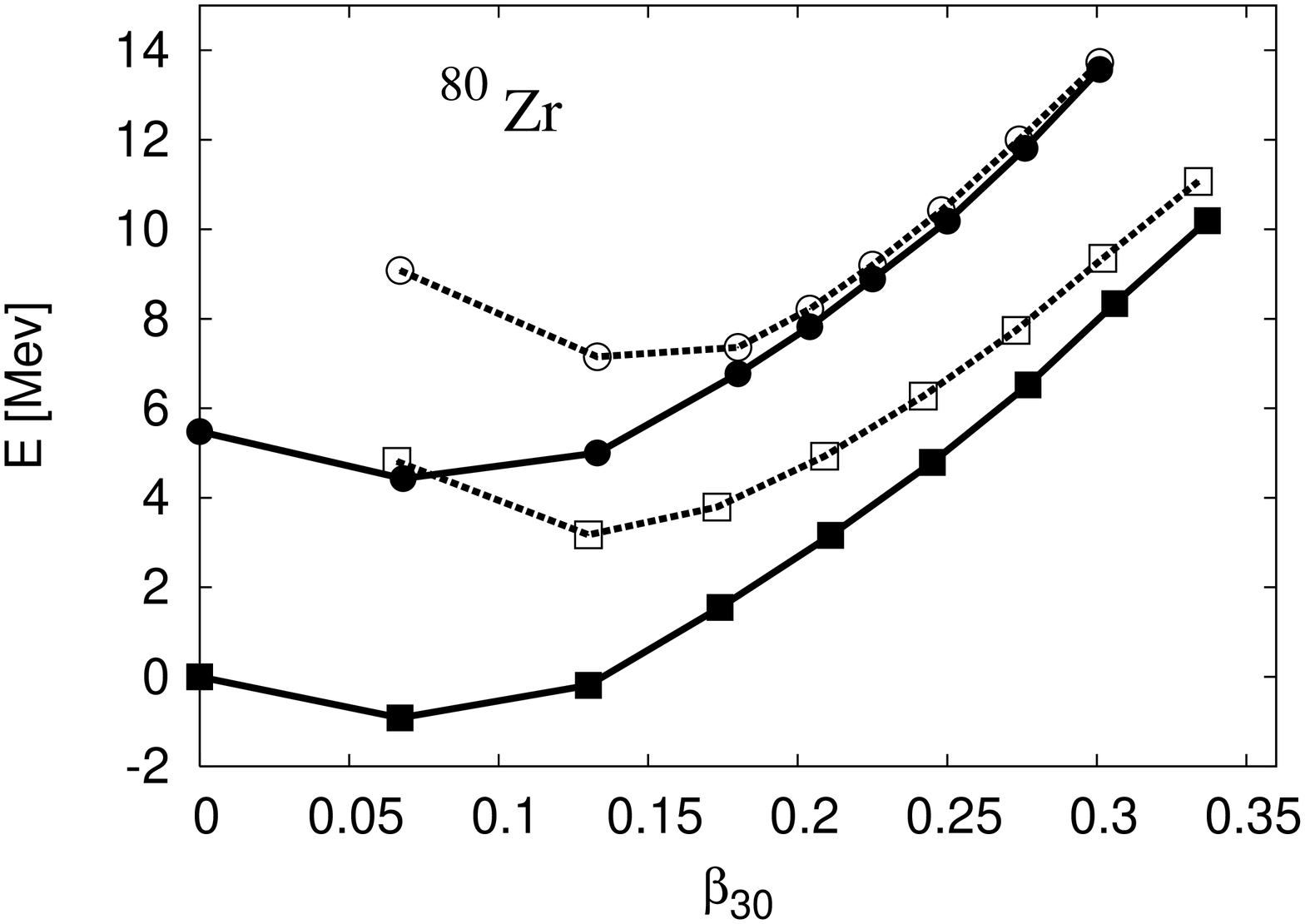}} &
      \resizebox{60mm}{!}{\includegraphics[angle=270]{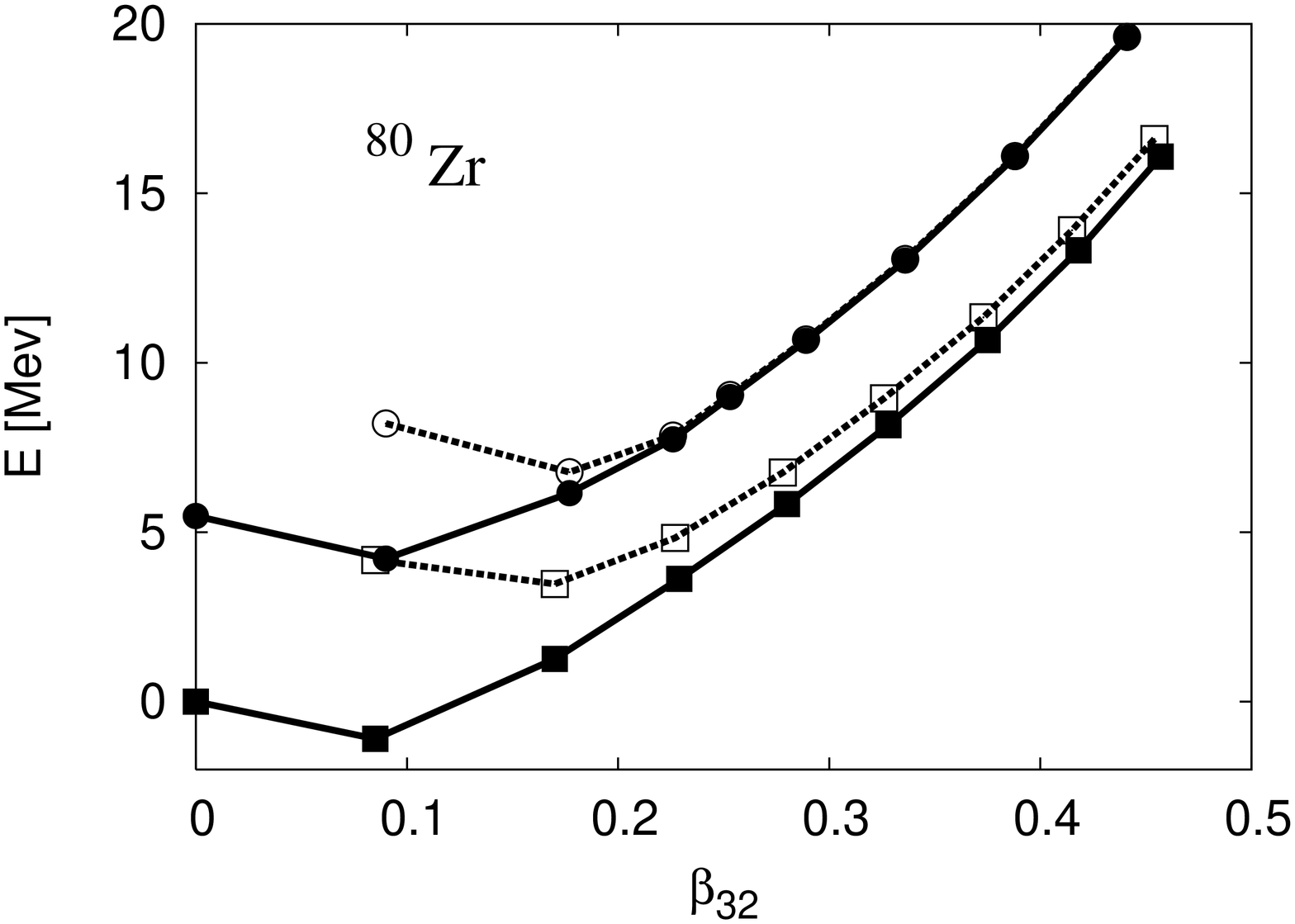}} \\
      \resizebox{60mm}{!}{\includegraphics[angle=270]{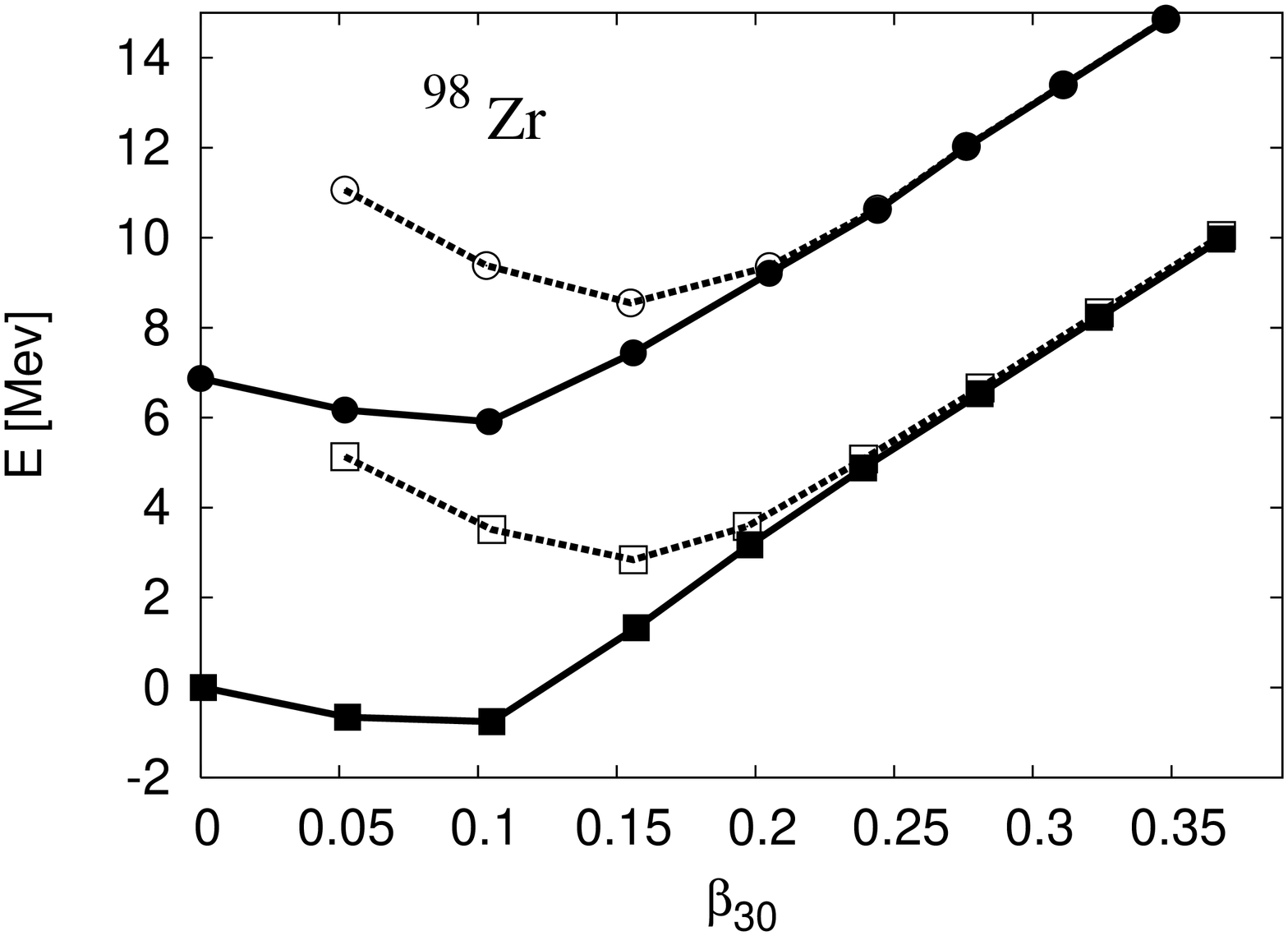}} &
      \resizebox{60mm}{!}{\includegraphics[angle=270]{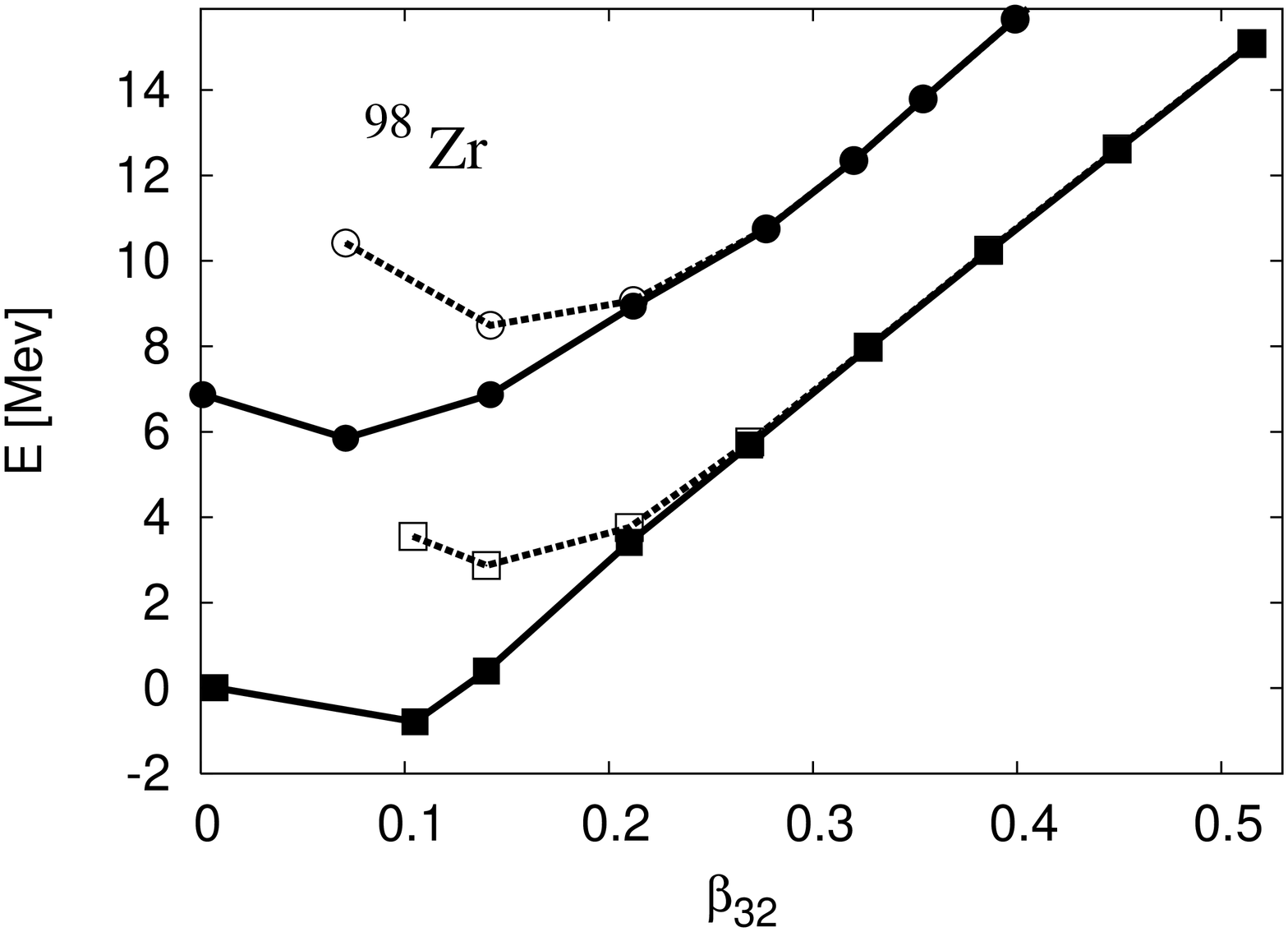}} \\
    \end{tabular}
    \caption{ \label{fig2} The parity projected mean-field energy as function of
shape parameters $\beta_{3\mu}$ obtained using the Sly4 force. Two
values of the pairing strength have been used: reproducing the
experimental gaps (circles) and with pairing gaps increased by a
factor 2 (squares). The solid (dotted) lines with filled (open)
symbols denote positive (negative) parity solutions.  }
   \end{center}
\end{figure}

It is instructive to investigate the sensitivity of the parity projected solutions
to the pairing strength. In the Fig.~\ref{fig2} we have shown the parity projected
energies for the SLy4 parametrization of the Skyrme force for normal and increased (twice)
strength of pairing correlations. Note that contrary to the behavior
of unprojected HFBCS energies, in this case the dependence of the energy
as function of octupole degrees of freedom practically remains unchanged.
In particular, the position of the energy minima for both negative and positive
parity states remains unaltered. Similar results have been obtained for
other Skyrme parametrizations.

\section{Generator coordinate method.}

\begin{figure}
  \begin{center}
    \begin{tabular}{cc}
      \resizebox{60mm}{!}{\includegraphics[angle=270]{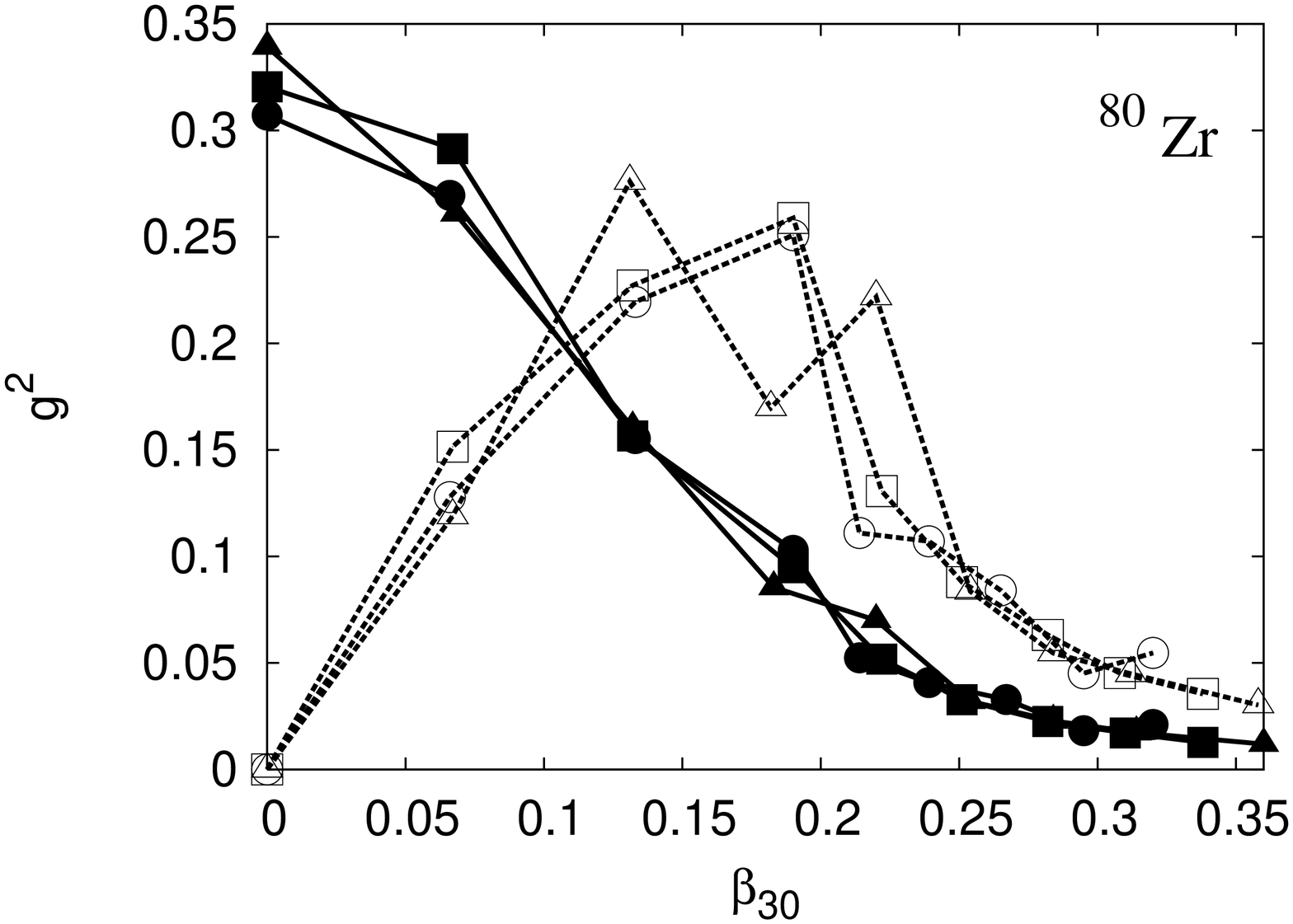}} &
      \resizebox{60mm}{!}{\includegraphics[angle=270]{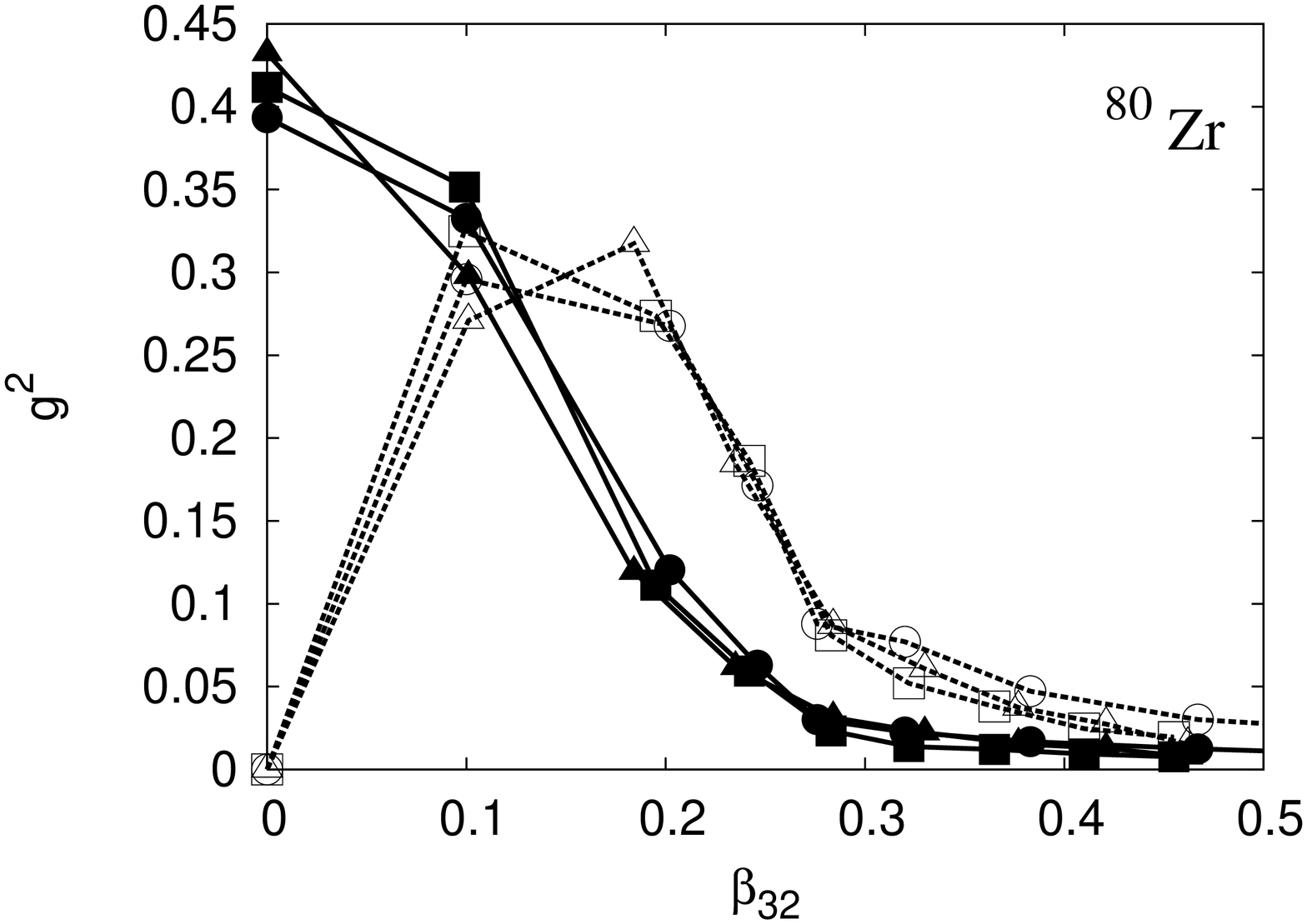}} \\
      \resizebox{60mm}{!}{\includegraphics[angle=270]{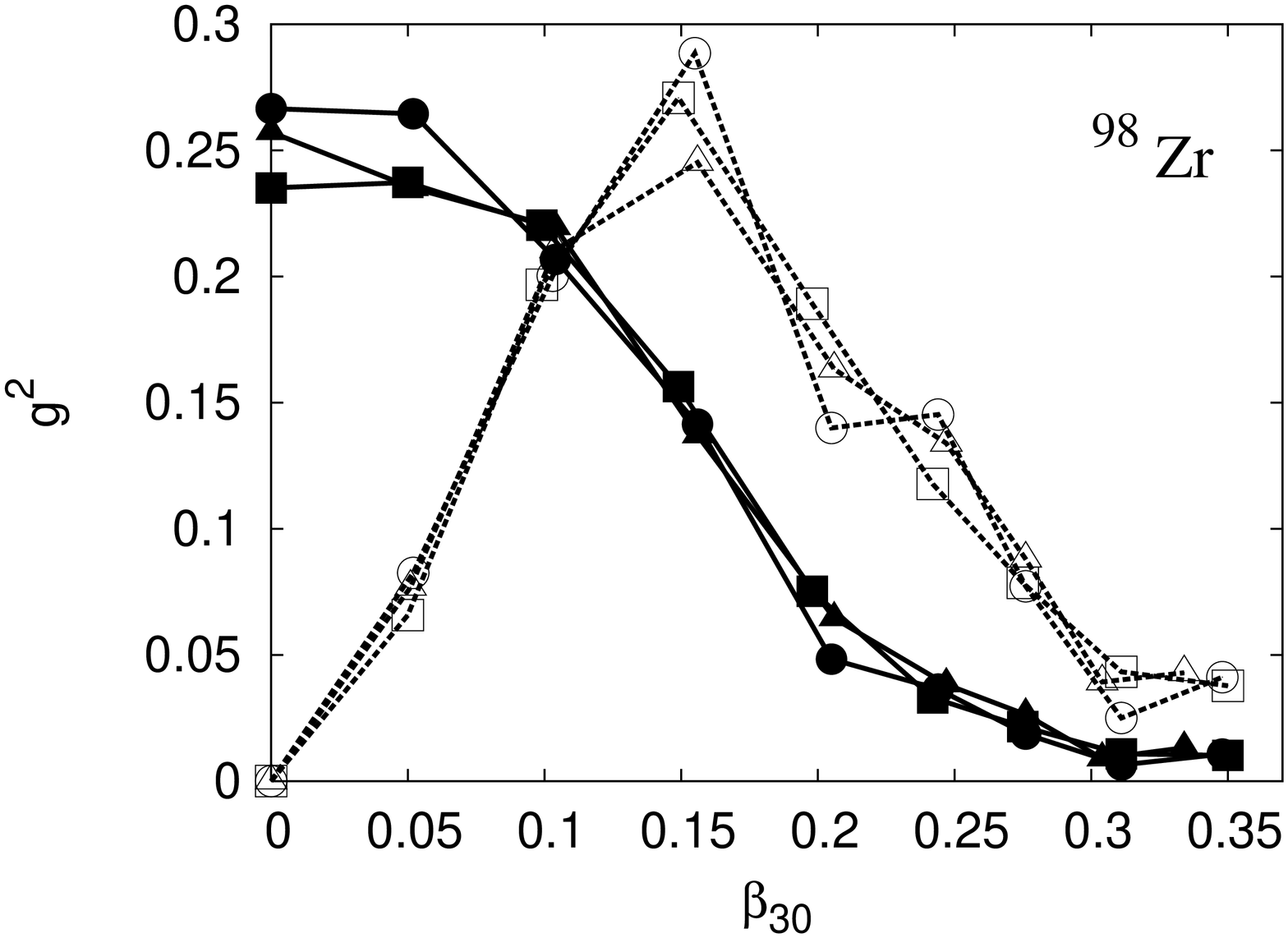}} &
      \resizebox{60mm}{!}{\includegraphics[angle=270]{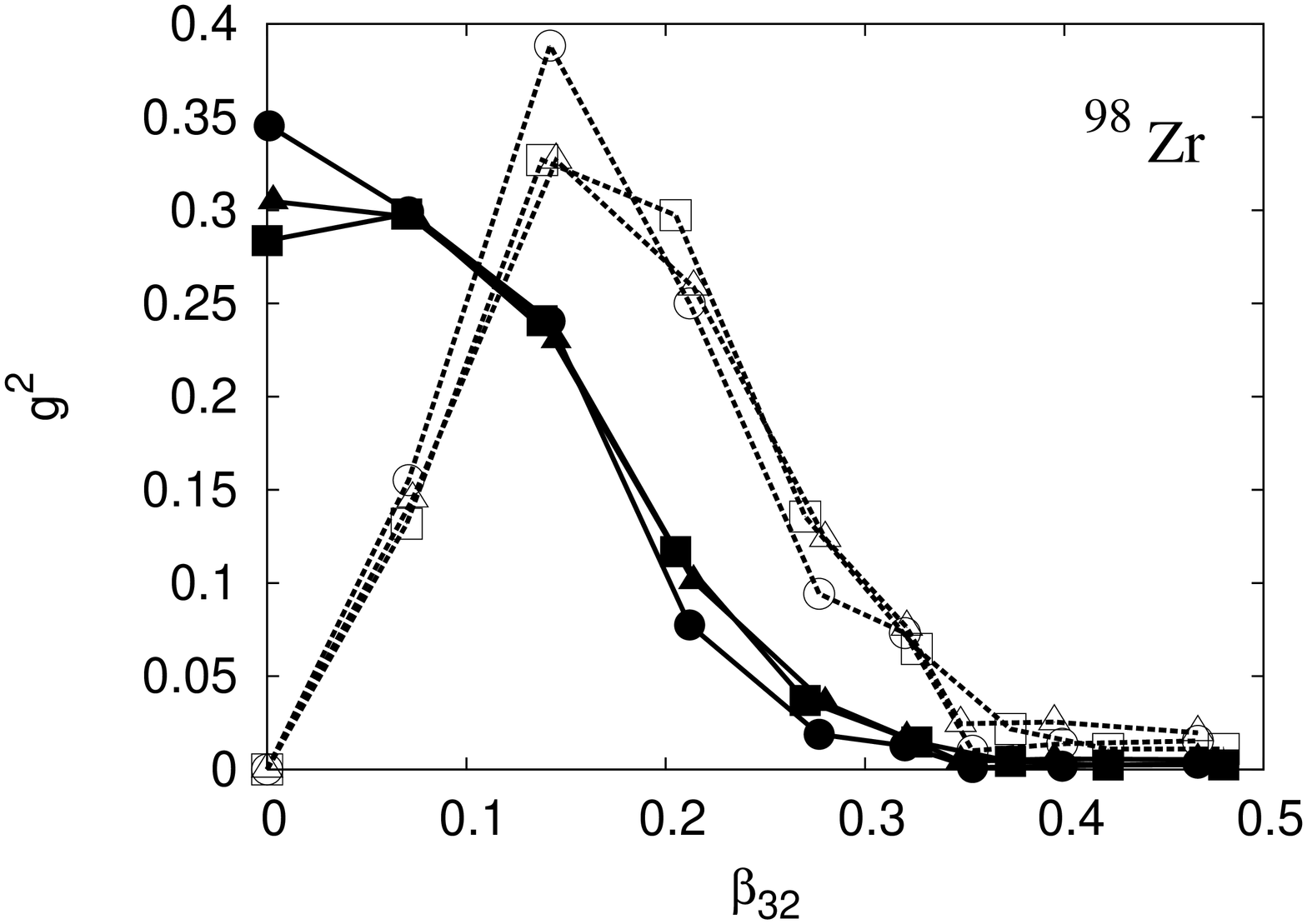}} \\
    \end{tabular}
    \caption{ \label{fig3} The square of the modulus of the
    collective wave function vs.
shape parameters $\beta_{3\mu}$ are plotted for three
parametrizations of the Skyrme force. The two figures at the
left-hand side show the energy as a function of the axial octupole
shape parameter $\beta_{30}$, whereas the two at the right-hand
side show the energy as function of $\beta_{32}$ (see text). The
symbols used for each curve are the same as in Fig. 1.}
  \end{center}
\end{figure}

\begin{figure}
  \begin{center}
    \begin{tabular}{cc}
      \resizebox{60mm}{!}{\includegraphics[angle=270]{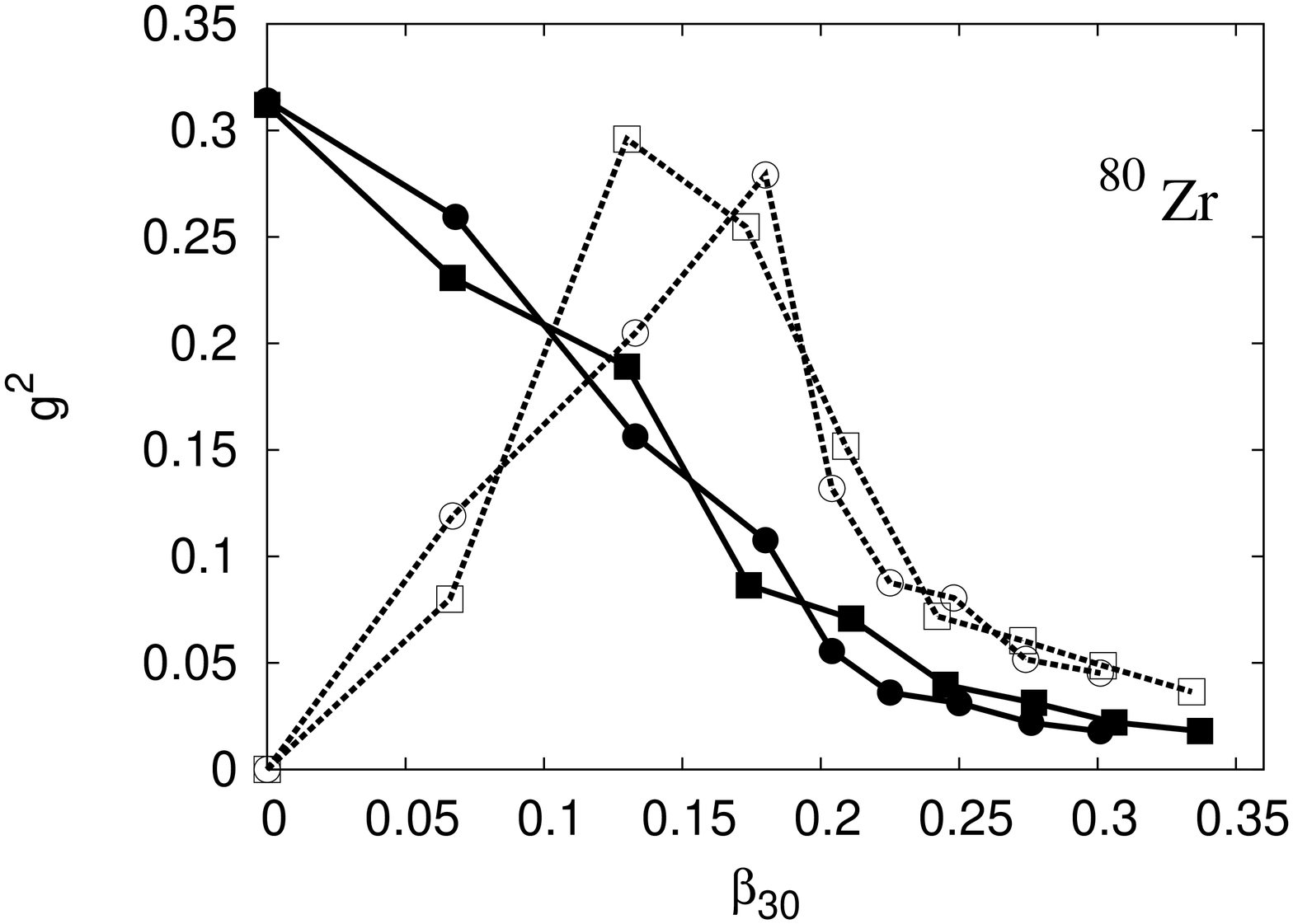}} &
      \resizebox{60mm}{!}{\includegraphics[angle=270]{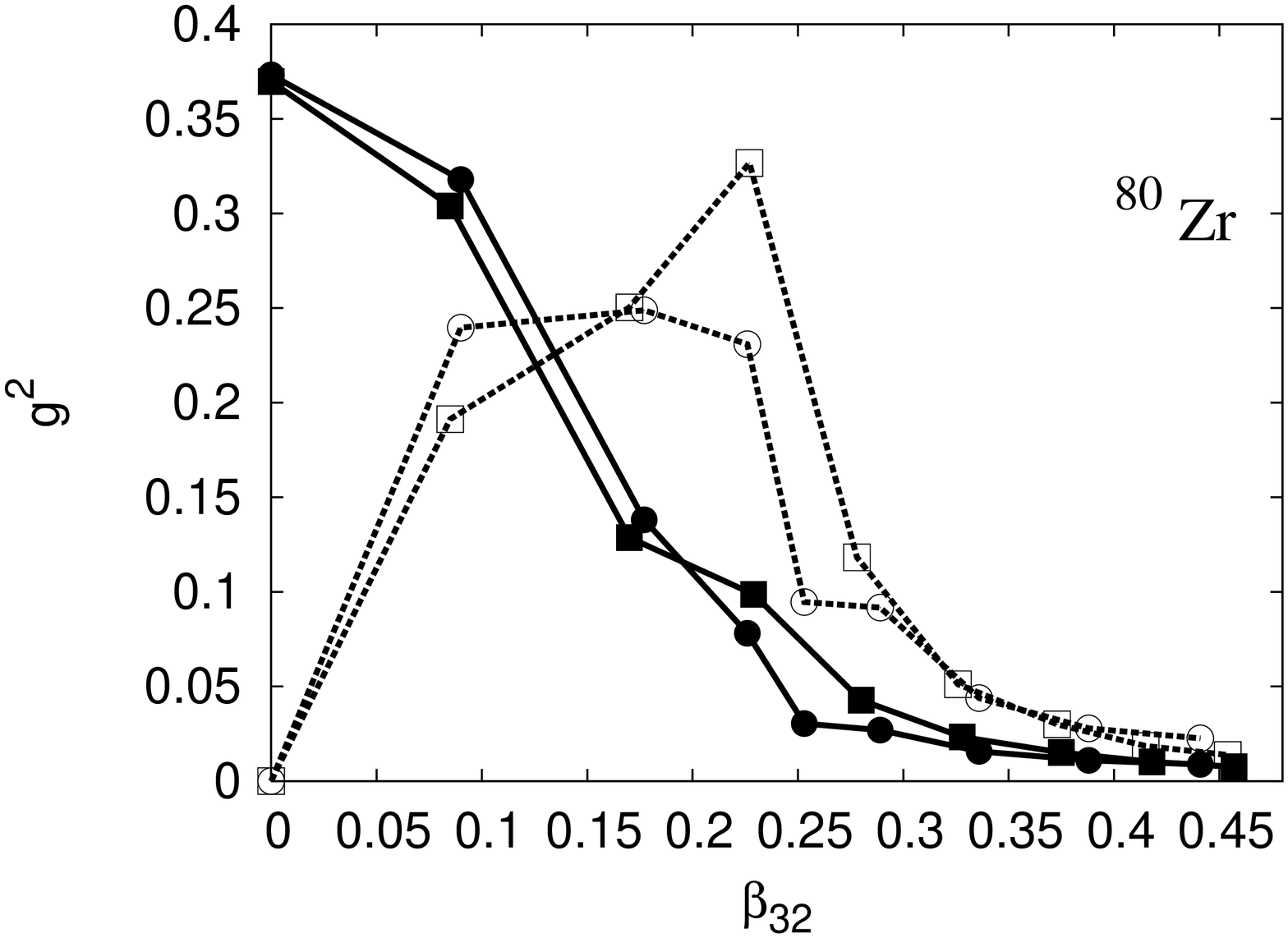}} \\
      \resizebox{60mm}{!}{\includegraphics[angle=270]{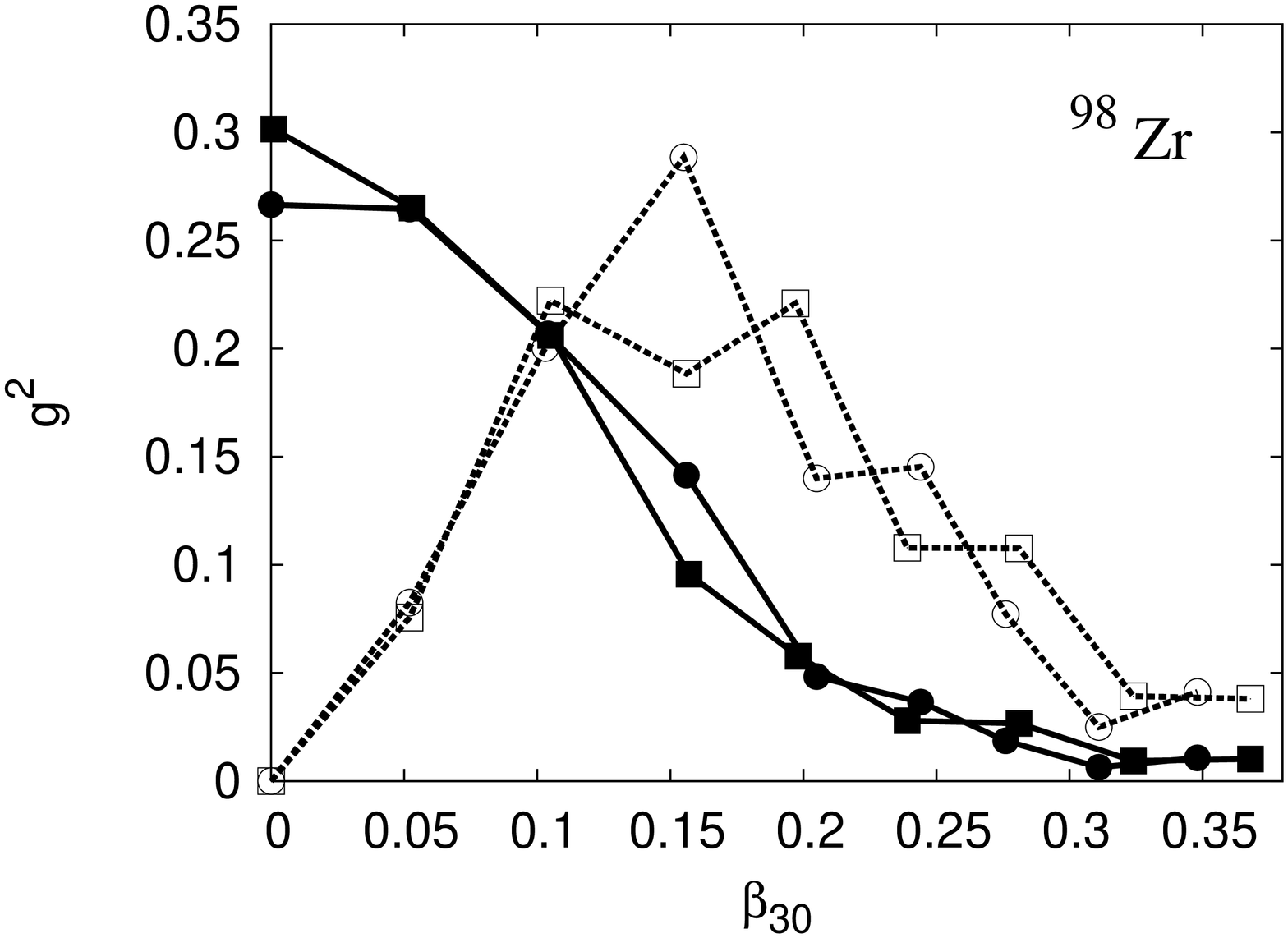}} &
      \resizebox{60mm}{!}{\includegraphics[angle=270]{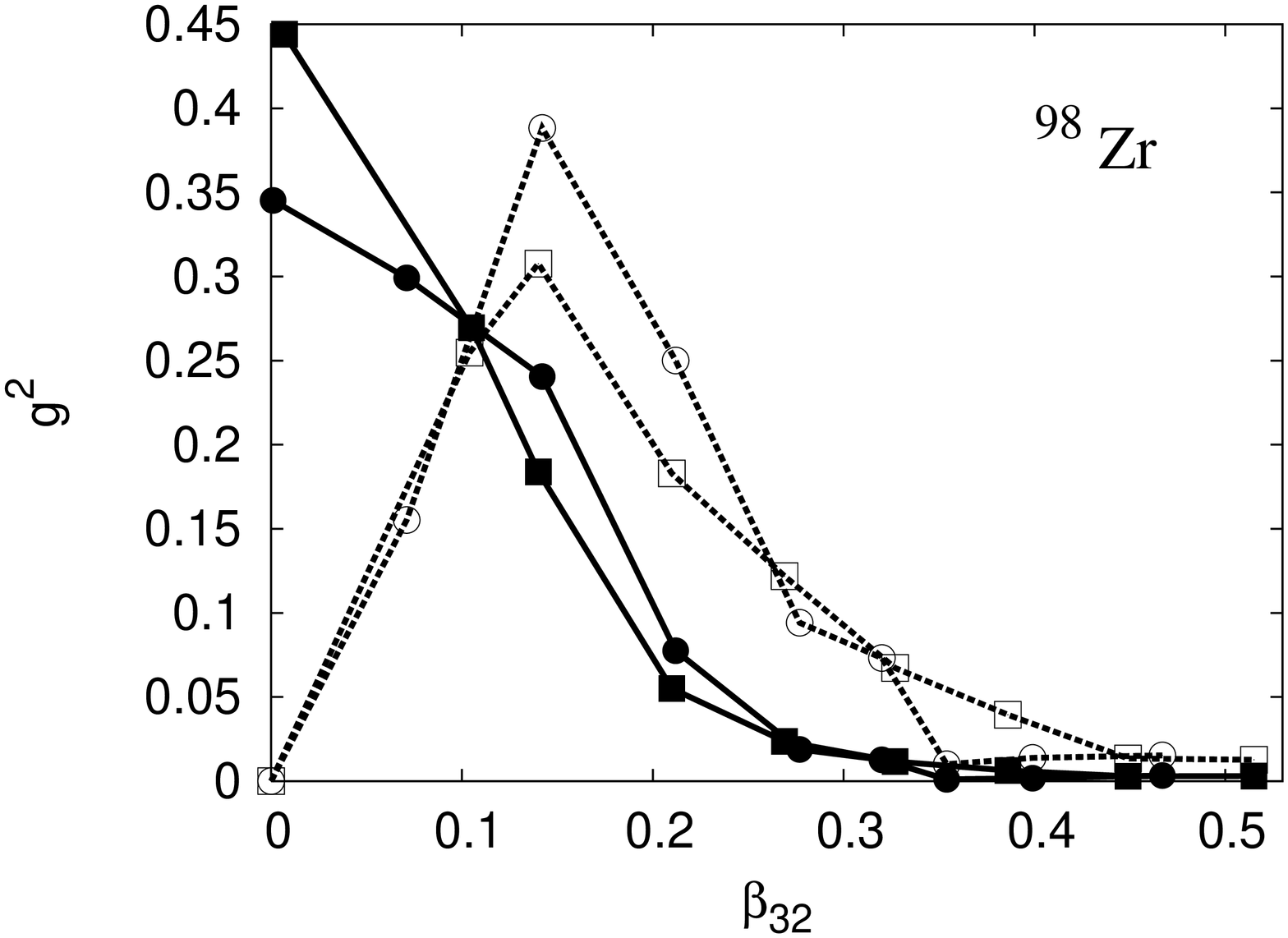}} \\
    \end{tabular}
    \caption{\label{fig4} The square of the modulus of the collective wave function vs.
shape parameters are plotted for the Sly4 force. Two values of
pairing strength have been considered: reproducing the
experimental pairing gaps (circles) and with pairing gaps
increased by a factor 2 (squares). The solid (dotted) lines with
filled (open) symbols denote positive (negative) parity
solutions.}
  \end{center}
\end{figure}

Mean-field results suggest that shape fluctuations
are important in this case. In order to quantify this effect we have applied
the generator coordinate method (GCM).  It allows to calculate
the correlation energies associated with shape fluctuations and
to determine the structure of the collective wave functions in terms of the
contributing mean-field configurations. Namely,
a collective wave function is constructed by mixing the mean-field
states corresponding to different values of the octupole
moment, after their projection on particle number and parity:
\beq
|\Psi\rangle = \int f(\beta_{3\mu}) \hat{P}_{(\pm, N,
Z)}|\phi(\beta_{3\mu})\rangle d \beta_{3\mu} .
\eeq
The coefficients $f(\beta_{3\mu})$ are determined by minimization
of the total energy of the collective wave function
$|\Psi\rangle$. In practice, the integral is
replaced by a discrete summation over $\beta_{3\mu}$, with a
number of points large enough to obtain results independent of the
discretization\cite{gcm,zber}. The discretized Hill-Wheeler (HW)
equation was solved separately for each collective coordinate
$Q_{30}$ and $Q_{32}$.

\begin{table}[pt]
\tbl{\label{tab:table2}Results of the GCM calculations for three Skyrme
      forces: SLy4, SIII and SkM$^{*}$. $E_{corr} (Q_{\lambda\mu})$ represents
      the correlation energy associated with shape fluctuation described
      by $Q_{\lambda\mu}$ multipole moment (see eq. \ref{CorrEner}). Similarly
      $E_{exc} (Q_{\lambda\mu})$ denotes the excitation energy of
      the first negative-parity collective state
      with respect to the first positive-parity state. All values are in $MeV$.}
{\begin{tabular}{@{}cccc@{}} \toprule
Nucleus & SLy4 & SIII & SkM$^{*}$  \\ \colrule
$^{80}$Zr& $E_{corr}(Q_{30})$ = 1.507 & 1.459 & 1.290   \\
         & $E_{corr}(Q_{32})$ = 1.567 & 1.499 & 1.338   \\
         & $E_{exc}(Q_{30})$  = 2.802 & 2.520 & 3.111   \\
         & $E_{exc}(Q_{32})$  = 2.425 & 2.220 & 2.881   \\
\hline
$^{98}$Zr& $E_{corr}(Q_{30})$ = 1.389 & 1.387 & 1.554 \\
         & $E_{corr}(Q_{32})$ = 1.400 & 1.485 & 1.564 \\
         & $E_{exc}(Q_{30})$  = 2.644 & 1.090 & 2.116 \\
         & $E_{exc}(Q_{32})$  = 2.498 & 0.784 & 1.776 \\ \botrule
\end{tabular}}
\end{table}

In the Fig.~\ref{fig3} the collective wave functions (related to
$f(\beta_{3\mu})$ by an integral transformation) are plotted
for three parametrizations of the Skyrme force.
One can see that all Skyrme forces give consistent results.
The wave functions are spread around
the minima of the projected mean-field energy curves, with a shape
typical of a vibration in a 1-dimensional energy well.
Note that there is no indication that any particular tetrahedral
configuration contribute the most to the wave function.
The large spread of the wave function confirms the importance of shape fluctuations.
It is reflected also in the large value of the correlation energies
which are listed in the table 1.
Correlation energies have been calculated from the prescription:
\beq
E_{corr} = E (N,Z,spher.) - E{^+},
\label{CorrEner}
\eeq
where $E (N,Z,spher.)$ is the energy of the
particle number projected spherical
configuration obtained in the HFBCS approach, and $E^{+}$ is the
lowest positive-parity energy obtained in the GCM.

Note also that the collective
wave functions for both axial octupole and tetrahedral
coordinates have a very similar shape. Indeed the calculated
correlation energies and excitation energies of the negative parity
state have very similar values, slightly favoring the tetrahedral
configuration.

In order to check the sensitivity of the GCM results on the
magnitude of pairing correlations, we have performed calculations
with pairing strengths that produce gaps larger or smaller
by a factor 2. Qualitatively, the GCM results are not affected
which can be seen in the Fig.~\ref{fig4}. The correlation energies vary by
$10-20$\%. A doubling of the pairing gap results in approximately
twice larger excitation energy for the negative parity states.

\section{Conclusions.}
\begin{itemize}
\item The existence of a stable tetrahedral deformation at the mean-field level is
      rather unlikely for $^{80}$Zr and $^{98}$Zr.
\item The parity projection induce a small tetrahedral (and also axial octupole)
      deformation which is relatively independent of the pairing strength.
\item Shape fluctuations play an important role and significantly contribute
      to the correlation energy.
\item Both axial octupole and tetrahedral states have very similar characteristics
      although the tetrahedral configuration is slightly more favored.
\end{itemize}
Summarizing we conclude that the existence of the tetrahedral deformation
is of the dynamic character.

\section*{Acknowledgements}
This work has been supported in part by the Polish Committee for Scientific
Research (KBN) under Contract No.~1~P03B~059~27, the
Foundation for Polish Science (FNP), the \mbox{PAI-P5-07} of the Belgian Office
for Scientific Policy.
Numerical calculations were performed at the Interdisciplinary Centre
for Mathematical and Computational Modelling (ICM) at Warsaw University.


\begin{thebibliography}{99}
\bibitem{hmx} I. Hamamoto, B. Mottelson, H. Xie, X. Z. Zhang,
              Z. Phys. {\bf D21} (1991) 163.
\bibitem{ld}  X. Li and J. Dudek,
              Phys. Rev. {\bf C49} (1994) R1250.
\bibitem{dgs} J. Dudek, A. G\'o\'zd\'z, N. Schunck, M. Mi\'skiewicz,
              Phys. Rev. Lett. {\bf 88} (2002) 252502.
              N. Schunck, J. Dudek, A. G\'o\'zd\'z and P.H. Regan,
              Phys. Rev. {\bf C69} (2004) 061305(R).
\bibitem{dud} J. Dudek, D. Curien, N. Dubray, J. Dobaczewski, V. Pangon, P. Olbratowski
              and N. Schunck, Phys. Rev. Lett. {\bf 97} (2006) 072501.
\bibitem{dgs2} J. Dudek, A. G\'o\'zd\'z and N. Schunck,
               Acta Phys.Polon. {\bf B34} (2003) 2491; N. Schunck, J. Dudek,
               Int. J. Mod. Phys. {\bf E13} (2004) 213.
\bibitem{zber} K. Zberecki, P. Magierski, P.-H. Heenen, N. Schunck, Phys. Rev. C (2006)
               {\em in press}, nucl-th/0604047.
\bibitem{bn}  P. Butler and W. Nazarewicz,
              Rev. Mod. Phys. {\bf 68} (1996) 349.
\bibitem{bfh05} P. Bonche, H. Flocard and P.-H. Heenen, Comp. Phys. Comm. {\bf 171} (2005) 49.
\bibitem{Hee91} P.-H. Heenen {\em et al.} Proc of Int. Workshop on Nuclear Structure Models,
                eds. R. Bengtsson, J. Draayer and W. Nazarewicz.  (ORNL 1992),
                World Scientific (Singapore), p 3.
\bibitem{ln} H.J. Lipkin,  Ann. of Phys. {\bf 9} (1960) 272.
\bibitem{dmn} J. Dobaczewski, P. Magierski, W. Nazarewicz, W. Satu{\l}a and
              Z. Szyma\'nski, Phys. Rev. {\bf C63} (2001) 024308.
\bibitem{yam} M. Yamagami, K. Matsuyanagi, M. Matsuo, Nucl. Phys. {\bf A693} (2001) 579.
\bibitem{tak} S. Takami, K. Yabana, M. Matsuo, Phys. Lett. {\bf B431}  (1998) 242.
\bibitem{polb} P. Olbratowski, J. Dobaczewski, P. Powa{\l}owski, M. Sadziak, K. Zberecki,
               Int. J. Mod. Phys. {\bf E13} (2006) 333.
\bibitem{shb93a} J. Skalski, P.-H. Heenen, P. Bonche, H. Flocard and J. Meyer,
                 Nucl. Phys. {\bf A551} (1993) 109.
\bibitem{gcm} M. Bender, P.-H. Heenen, and P.-G. Reinhard,
              Rev. Mod. Phys. {\bf 75} (2003) 121.
\end{thebibliography}
\end{document}